\documentclass[12pt]{article}
\usepackage{a4wide}
\usepackage{amssymb, amsmath}
\usepackage[normalem]{ulem}
\usepackage{graphicx}
\usepackage{cite}
\usepackage[utf8]{inputenc}
\usepackage{subfigure}
\usepackage[colorlinks]{hyperref}
\usepackage[usenames,dvipsnames]{color}
\hypersetup{
     breaklinks=true,
    pdfstartview={FitH},    
    colorlinks=true,       
    linkcolor=blue,          
    citecolor=red,        
    filecolor=magenta,      
    urlcolor=blue,           
    anchorcolor=green,      
    linktocpage=true
}

\begin{document}
{\renewcommand{\thefootnote}{\fnsymbol{footnote}}
		
\begin{center}
{\LARGE Dark energy beyond quintessence: \\
	Constraints from the swampland} 
\vspace{1.5em}

Suddhasattwa Brahma$^{1}$\footnote{e-mail address: {\tt suddhasattwa.brahma@gmail.com}} and Md. Wali Hossain$^{1}$\footnote{e-mail address: {\tt wali.hossain@apctp.org}}
\\
\vspace{0.5em}
$^1$ Asia Pacific Center for Theoretical Physics,\\
Pohang 37673, Korea\\
\vspace{1.5em}
\end{center}
}
	
\setcounter{footnote}{0}

\newcommand{\bea}{\begin{eqnarray}}
\newcommand{\eea}{\end{eqnarray}}
\renewcommand{\d}{{\mathrm{d}}}
\renewcommand{\[}{\left[}
\renewcommand{\]}{\right]}
\renewcommand{\(}{\left(}
\renewcommand{\)}{\right)}
\newcommand{\nn}{\nonumber}
\newcommand{\Mpl}{M_{\textrm{Pl}}}
\def\H{\mathrm{H}}
\def\V{\mathrm{V}}
\def\e{\mathrm{e}}
\def\be{\begin{equation}}
\def\ee{\end{equation}}

\def\al{\alpha}
\def\bet{\beta}
\def\gam{\gamma}
\def\om{\omega}
\def\Om{\Omega}
\def\sig{\sigma}
\def\Lam{\Lambda}
\def\lam{\lambda}
\def\ep{\epsilon}
\def\ups{\upsilon}
\def\vep{\varepsilon}
\def\S{\mathcal{S}}
\def\doi{http://doi.org}
\def\arxiv{http://arxiv.org/abs}
\def\d{\mathrm{d}}
\def\g{\mathrm{g}}
\def\m{\mathrm{m}}
\def\r{\mathrm{r}}

\begin{abstract}
\noindent The string theory swampland proposes that there is no UV-completion for an effective field theory with an exact (metastable) de Sitter vacua, thereby ruling out standard $\Lambda$CDM cosmology if the conjecture is taken seriously. The swampland criteria have also been shown to be in sharp tension with quintessence models under current and forthcoming observational bounds. As a logical next step, we introduce higher derivative self-interactions in the low-energy effective Lagrangian and show that one can satisfy observational constraints as well as the swampland criteria for some specific models. In particular, the cubic Galileon term, in the presence of an exponential potential, is examined to demonstrate that parts of the Horndeski parameter space survives the swampland and leads to viable cosmological histories.
\end{abstract}

\section{Introduction}
Precision cosmological observations have firmly established the evolution history of our universe, laying down the existence of a current period of exponential expansion. However, the theoretical seeds for this late-time acceleration remain a source of considerable debate. Occam's razor dictates that the simplest explanation lies in assuming a positive cosmological constant leading to the standard $\Lambda$CDM paradigm. Already at this level, there are perplexing questions regarding the `naturalness' of the small value of $\Lambda$ as observed in our universe, and its compatibility with our current understanding of vacuum energy density. However, even allowing for this observed value of the cosmological constant as a free parameter in standard cosmology, it is interesting to ask if there exists other theoretical paths towards constraining the $\Lambda$CDM model. 

On the other hand, the ultraviolet (UV) completion of general relativity (GR), or the problem of quantum gravity, remains one of the most interesting challenges in theoretical physics today. In the most exciting case, peculiarities of the UV theory, whatever it might be, can perhaps leave its signatures on GR -- the empirically verified low energy effective field theory (EFT) of gravity. Recently, it has been proposed that there is a swampland of inconsistent EFTs, which are incompatible with string theory, unless they satisfy some constraints \cite{Ooguri:2006in, Obied:2018sgi, Agrawal:2018own, Ooguri:2018wrx}
\begin{itemize}
	\item[(S1):] The range traversed by any of the scalar field, $\pi$, has the upper bound, $|\Delta \pi|/\Mpl < \Delta \sim \mathcal{O}(1)$, AND
	\item[(S2):] The gradient of the scalar field potential, $V(\pi)$, is bounded from below, $\Mpl\, |V'|/V > c \sim \mathcal{O}(1)$,\, OR \\
	The potential has unstable directions with large curvature, \textit{i.e.} $-\Mpl^2\, \text{min}(V'')/V > \tilde{c} \sim \mathcal{O}(1)$,
\end{itemize}
where $\Delta$, $c$ and $\tilde{c}$ are positive and $\Mpl$ is the reduced Planck mass. Specifically, the second condition above implies that there are no metastable de-Sitter (dS) solutions in EFTs emerging from string theory. This conjecture, referred from now on as the `dS constraint', has led to a new guideline for constraining cosmological models \cite{Agrawal:2018own,Blumenhagen:2017cxt,Achucarro:2018vey,Garg:2018reu,Dias:2018ngv,Kehagias:2018uem,Lehners:2018vgi,Denef:2018etk,Colgain:2018wgk,Heisenberg:2018yae,Akrami:2018ylq,Heisenberg:2018rdu,Kinney:2018nny,Brahma:2018hrd,Das:2018hqy,Wang:2018duq,Danielsson:2018qpa,Han:2018yrk,Visinelli:2018utg,Matsui:2018xwa,Hamaguchi:2018vtv,Das:2018rpg,Lin:2018kjm,Kawasaki:2018daf,Dimopoulos:2018upl,Motaharfar:2018zyb,Ashoorioon:2018sqb,Wang:2018kly,Fukuda:2018haz,Garg:2018zdg,Park:2018fuj,Lin:2018rnx,Schimmrigk:2018gch,Agrawal:2018rcg,Yi:2018dhl,Heckman:2018mxl,Elizalde:2018dvw,Cheong:2018udx,Holman:2018inr,Acharya:2018deu,Herdeiro:2018hfp,Kinney:2018kew,Montero:2018fns,Lin:2018edm,Cai:2018ebs,Heckman:2019dsj,Kamali:2019hgv,Haque:2019prw,Andriot:2018wzk,Raveri:2018ddi,Murayama:2018lie,Odintsov:2018zai,Heisenberg:2019qxz} and shall be the main focus of our work.

The dS constraint clearly rules out standard $\Lambda$CDM cosmology, our simplest proposal for late-time acceleration, due to the unavailability of any local dS extrema in the effective potential. As an alternative, it has been proposed that the current era of accelerated expansion be explained by models of quintessence, i.e. by assuming a scalar field beyond the Standard Model. (In a similar manner, the same swampland conjectures have also put appreciable pressure on the simplest models of vanilla single-field inflation -- an era of accelerated expansion in the early-universe -- thereby necessitating the introduction of more complicated models, \textit{e.g.} \cite{Brahma:2018hrd,Kinney:2018nny,Achucarro:2018vey,Garg:2018reu,Kehagias:2018uem,Blumenhagen:2017cxt,Das:2018rpg,Motaharfar:2018zyb,Lin:2018kjm,Dias:2018ngv}.) On the bright side, such quintessence models can easily be embedded in string theory through slowly rolling moduli fields (see, for instance  \cite{Cicoli:2012tz,Gupta:2011yj, Panda:2010uq, Hellerman:2001yi, Choi:1999xn}), thereafter suitably imposing the swampland conjectures on them. However, it was also noted that current cosmological observations put a strict bound on the $c$, which for the least constrained exponential potential, is put at $c<0.6$ at $1\sigma$ level \cite{Heisenberg:2018yae,Heisenberg:2018rdu,Fukuda:2018haz,Wang:2018duq}. Therefore, one is forced to consider modifications to GR, which typically contain an additional scalar field, as a necessary step for realizing models of evolving dark-energy in order to satisfy the dS constraint coming out of the swampland conjectures.

Our main motivation for this work is to show that this bound can be significantly relaxed when considering models which go beyond quintessence; in particular, by considering higher derivative terms in the action as is common for Galileon theories \cite{Nicolis:2008in,Chow:2009fm,Silva:2009km,DeFelice:2010pv,DeFelice:2010nf,Deffayet:2009wt}. The most general class of such theories with a scalar field, in the presence of derivative interactions, is represented by the Horndeski Lagrangian \cite{Horndeski:1974wa,Deffayet:2009mn,Deffayet:2013lga,Kase:2018aps}, the viability of which after imposing the swampland constraints has also been studied recently \cite{Heisenberg:2019qxz}. Although we do not study the most general class of full Horndeski theories, we point out, through our simpler example, that there are parts of the parameter space which naturally fit the late-time data much better than quintessence models, even after taking the swampland conjectures into account, once derivative interactions are turned on. Therefore, to reemphasize our main result, there are parts of the parameter space of Horndeski Lagrangian which are not in the swampland and can be efficiently used for model-building in explaining our cosmic history. We devote the rest of this paper to flesh out the details of our simple model which shall act as a proof of principle in support of this claim. 

\section{The model: cubic Galileon terms}
In our model, the late-time acceleration of the universe shall be explained through the dynamics of an effective scalar field -- the Galileon  $\pi$. We consider the following cubic Galileon action with a potential term $V(\pi)$ \cite{Ali:2012cv,Hossain:2012qm,Hossain:2017ica,Dinda:2017lpz}, a sub-class of the most general scalar-tensor (Horndeski) Lagrangian
\begin{equation}
	\S=\int \d^4x\sqrt{-\g}\Bigl [\frac{\Mpl^2}{2} R-\frac{1}{2}(\nabla \pi)^2\Bigl(1+\frac{\al}{M^3}\Box \pi\Bigr) - V(\pi) \Bigr]+ \S_\m+\S_\r \, ,
	\label{eq:action}
\end{equation}
where $\Mpl=1/\sqrt{8\pi G}$ is the reduced Planck mass, $M$ is an energy scale, $\al$ is a dimensionless constant. $\S_\m$ and $\S_\r$ are the matter and radiation action. In the above, one can rescale the parameter $\al$ to replace $M$ by $\Mpl$. It is also straightforward to see that setting $\alpha=0$ gets us back to the standard quintessence action.

When potential $V(\pi)$ is linear, the scalar field action preserves the Galilean shift symmetry $\pi\to\pi+b_\mu x^\mu+c$, where $b_\mu$ and $c$ are constants. However, an exponential potential, which is what we shall consider in this work,  results in the breaking of this shift symmetry. Note that our primary aim is to show how the swampland conjectures can easily be satisfied when including non-canonical higher-derivative terms in the action, going beyond quintessence models. As it is, quantum corrections can lead to the appearance of higher derivative terms when the non-renormalizable theorem is violated \cite{Pirtskhalava:2015nla}, when the Galileon symmetry is not \textit{weakly-broken}. Our choice of the exponential potential is purely phenomenological in spirit and takes this specific form to facilitate comparison with quintessence \cite{Heisenberg:2018yae} and some classes of Horndeski theories \cite{Heisenberg:2019qxz}, when combined with the swampland conjectures\footnote{However, independently from these considerations, one needs to add a potential term so as to get ghost-free late-time acceleration in models of cubic Galileon \cite{Ali:2010gr,Gannouji:2010au}.}. 

Apart from the potential term, one also needs to make a choice for the higher derivative terms to keep in the action. While the quartic/quintic terms in the covariant Galileon formalism has been made nonviable since the detection of multi-messenger gravitational wave astronomy \cite{Ezquiaga:2017ekz,Sakstein:2017xjx}\footnote{For earlier work, see \cite{Lombriser:2015sxa,McManus:2016kxu}.}, the cubic term is highly disfavored by Integrated Sachs-Wolfe effect (ISW) measurements \cite{Renk:2017rzu}. However, generalizations of the cubic Galileon models have been shown to be compatible with the ISW data \cite{Kimura:2011td}. Keeping these in mind, we shall only consider the cubic Galileon term, $(\nabla\pi)^2\, \Box\pi$, in our action. We remind the reader that not only has this term not been ruled out by gravitational wave detection, neither it been shown to be in conflict with the ISW data in the presence of a potential term \cite{Kase:2018aps}, as is the case here. Finally, and most importantly, we emphasize that this specific model has been chosen only to provide a concrete example to illustrate that parts of the parameter space of general Horndeski theories remain viable even after the imposition of the swampland conjectures.

Varying the action (\ref{eq:action}) with respect to (w.r.t.) the metric $\g_{\mu\nu}$ gives the Einstein equation  
\begin{equation}
	\Mpl^2 G_{\mu\nu}= T_{(\m)\mu\nu}+T_{(\r)\mu\nu}+T_{(\pi)\mu\nu} \, ,
	\label{eq:ee}
\end{equation}
where subscripts $m$, $r$ and $\pi$ represent matter, radiation and the scalar field respectively and
\begin{align}
	T_{(\pi)\mu\nu}=& \pi_{;\mu}\pi_{;\nu}-\frac{1}{2}\g_{\mu\nu}(\nabla\pi)^2 -\g_{\mu\nu}V(\pi)+\frac{\alpha}{M^3} \Bigl[\pi_{,\mu}\pi_{;\nu}\Box\pi+\g_{\mu\nu}\pi_{;\lambda}\pi^{;\lambda\rho}\pi_{;\rho} \nn \\ & - \pi^{;\rho}\(\pi_{;\mu}\pi_{;\nu \rho}+\pi_{;\nu}\pi_{;\mu \rho}\)\Bigr] \, ,
	\label{eq:emt_phi}
\end{align}
and varying w.r.t. the scalar field $\pi$ gives the equation of motion of the scalar field 
\begin{equation}
	\Box \pi+\frac{\alpha}{M^3}\Bigl[(\Box\pi)^2-\pi_{;\mu\nu}\pi^{;\mu\nu}-R^{\mu\nu}\pi_{;\mu}\pi_{;\nu}\Bigr]-V'(\pi)=0 \, ,
	\label{eq:eom_phi}
\end{equation}
where $'$ denotes the derivative w.r.t. $\pi$.


\section{Cosmological dynamics constrained by the Swampland}

In a spatially flat  Friedmann-Lema\^{i}tre-Robertson-Walker (FLRW) background, the Friedmann equations take the form
\begin{eqnarray}
	3M_{\rm{pl}}^2H^2 &=&\rho_\m+\rho_\r + \rho_\pi\,,\label{Friedmann1}\\
	M_{\rm{pl}}^2(2\dot H + 3H^2)&=&-\frac{\rho_\r}{3} - P_\pi\,,\label{Friedmann2}
\end{eqnarray}
where
\begin{eqnarray}
	\rho_\pi &=& \frac{\dot{\pi}^2}{2}\Bigl(1-\frac{6\alpha}{M^3} H\dot{\pi}\Bigr) + V{(\pi)} \, ,
	\label{eq:rhopi}\\
	P_\pi &=& \frac{\dot{\pi}^2}{2} \Bigl(1+\frac{2\alpha}{M^3}\ddot{\pi}\Bigr) - V(\pi) \, .
	\label{eq:ppi}
\end{eqnarray}

and the equation of motion for the scalar field reads
\begin{eqnarray}
	\ddot{\pi}+ 3H\dot{\pi}- \frac{3\alpha}{M^3} \dot{\pi}\Bigl(3H^2\dot{\pi} + \dot{H}\dot{\pi} + 2H\ddot{\pi}\Bigr)+ V'(\pi)=0 \, ,
\end{eqnarray}
where $\rho_\m$ and $\rho_\r$ are the energy densities of non-relativistic matter ($P_\m = 0$) and radiation ($P_\r = \rho_\r/3$) respectively.

To examine the background cosmological dynamics, let us define the following dimensionless variables \cite{Ali:2012cv,Hossain:2012qm}
\begin{eqnarray}
	x &=& \frac{\dot{\pi}}{\sqrt{6}H M_{\rm{pl}}}\,, 
	\label{eq:x1}\\
	y &=& \frac{\sqrt{V}}{\sqrt{3} H M_{\rm{pl}}}\,,
	\label{eq:y1}\\
	\epsilon &=& -6\frac{\alpha}{M^3}H\dot \pi\,,
	\label{eq:ep1}\\
	\Omega_\r &=& \frac{\rho_\r}{3 \Mpl^2 H^2}\,,
	\label{eq:omr1}\\
	\lambda &=& -M_{\rm{pl}}\frac{V'}{V},
	\label{eq:lam1}
\end{eqnarray}
and the equation-of-state (EoS) parameters,
\begin{eqnarray}
	w_{\rm eff} &=& -1-\frac{2}{3}\frac{\dot H}{H^2} = \frac{3 x^2 \(4+8\ep+\ep^2\)-2\sqrt{6}xy^2\ep\lam-4(1+\ep)\(3y^2-\Om_\r\)}{3\(4+4\ep+x^2\ep^2\)} \, , \\
	w_\pi &=& \frac{P_\pi}{\rho_\pi} = -\frac{12 y^2 (1+\epsilon )+2 \sqrt{6} x y^2 \ep\lam-x^2 \(12+24\ep+\ep^2(3-\Om_\r)\)}{3\left(4+4 \epsilon +x^2 \epsilon ^2\right) \left(y^2+x^2 (1+\epsilon )\right)} \, ,
\end{eqnarray}
where $w_{\rm eff}$ and $w_\pi$ are the effective and scalar field EoS.

The evolution equations of the dimensionless variables form the following autonomous system 
\begin{align}
	\frac{{\rm d}x}{{\rm d}N}&=x\Bigl(\frac{\ddot{\pi}}{H\dot{\pi}}-\frac{\dot H}{H^2}\Bigr)
	\label{eq:x}\\
	\frac{{\rm d}y}{{\rm d}N}&=-y \Bigl(\sqrt{\frac{3}{2}}\lambda x+\frac{\dot H}{H^2}\Bigr)
	\label{eq:y}\\
	\frac{{\rm d}\epsilon}{{\rm d}N}&=\epsilon \Bigl(\frac{\ddot{\pi}}{H\dot{\pi}}+\frac{\dot H}{H^2}\Bigr)
	\label{eq:ep}\\
	\frac{{\rm d}\Omega_r}{{\rm d}N}&=-2\Omega_r\Bigl(2+\frac{\dot H}{H^2}\Bigr)
	\label{eq:omr}\\
	\frac{{\rm d}\lambda}{{\rm d}N}&=\sqrt{6}x\lambda^2(1-\Gamma)
\end{align}
where $N=\text{ln}\, a$ is the number of $e$-folds. Here $\Gamma=VV_{,\pi\pi}/V_{,\pi}^2$ whereas, using Eqs.~\eqref{Friedmann1}-\eqref{Friedmann2}, 
\begin{align}
	\frac{\dot H}{H^2}&=\frac{2(1+\epsilon)(-3+3y^2-\Omega_r)-3x^2(2+4\epsilon+\epsilon^2)+\sqrt{6}x\epsilon y^2\lambda}{4+4\epsilon+x^2\epsilon^2} \, ,\\
	\frac{\ddot{\pi}}{H\dot{\pi}}&=\frac{3x^3\epsilon-x\Bigl(12+\epsilon (3+3y^2-\Omega_r)\Bigr)+2\sqrt{6}y^2\lambda}{x(4+4\epsilon+x^2\epsilon^2)} \, .
\end{align}

For our choice of an exponential potential with a constant slope $\lambda$, $V(\pi) = e^{-\lambda \pi/\Mpl}$, implies $\Gamma=1$. In this case, the autonomous system can be reduced to the Eqs.~(\ref{eq:x})-(\ref{eq:omr}). Defining the matter density $\Omega_\m := \rho_\m/(3\Mpl^2 H^2)$, we can rewrite the constraint equation as $\Omega_\m+\Omega_\r+\Omega_\pi=1$ where 
\begin{equation}
	\Omega_\pi=x^2(1+\epsilon)+y^2 \, .
	\label{eq:ompi}
\end{equation}

\subsection{Fixed point analysis}

It is straightforward to see from the Eqs.~(\ref{eq:rhopi}), (\ref{eq:ppi}) and (\ref{eq:ep1}) that $\epsilon$ parameterizes corrections in the dynamics due to the cubic Galileon term and setting $\epsilon=0$ gives the autonomous system for the standard quintessence scenario. Let us first briefly discuss the fixed point analysis for the cubic Galileon model with an exponential potential. We will also review the late time cosmological dynamics of the scalar field for an exponential potential. As shall be emphasized later on, we note that there exist several differences in the structure of critical points on our phase space as compared to the Horndeski model studied in \cite{Heisenberg:2019qxz,Kase:2015zva}

The physical fixed points are listed in the Table~\ref{tab:fp}. The fixed points and their nature of stability is similar to the quintessence field with exponential potential. However, it is crucial to note that the inclusion of the variable $\ep$ reduces the number of the fixed points which one has for the quintessence field alone ($\ep=0$), including an exponential potential. For instance, the fixed point characterized by $\left(x,y,\Omega_\r\right) = \left(0,0,1\right)$ is absent in our case. Moreover, it also affects the character of some the points (see $\rm C_\pm$) and makes them behave strictly as saddle points whereas, for the quintessence case, they can act as an unstable nodes as well \cite{Copeland:1997et}. Interestingly, as can be seen from the table below, $\ep$ flows to zero for all the critical points in our model on a four-dimensional phase space. 

\begin{table*}[ht]
	\begin{center}
		\resizebox{\textwidth}{!}{%
			\begin{tabular}{|c|c|c|c|c|c|c|c|c|c|c|}\hline \hline
				Pts. & $x$ & $y$ & $\ep$ &  $\Om_\r$ & Existence & Stability & $\Omega_\m$ & $\Omega_\pi$ & $w_\pi$ & $w_{\rm eff}$ \\ 
				\hline\hline 
				$\rm A_\pm$ & $2\sqrt{2}/(\sqrt{3}\lam)$ & $\pm 2/(\sqrt{3}\lam)$ & 0 & $1-4/\lam^2$ & $\lam^2>4$ &Saddle point & 0 & $4/\lam^2$ & 1/3 & 1/3 \\
				\hline
				$\rm B_\pm$ & $\sqrt{3}/(\sqrt{2}\lam)$ & $\pm \sqrt{3}/(\sqrt{2}\lam)$ & 0 & 0 & $\lam^2>3$ & Saddle Point for & $1-3/\lam^2$ & $3/\lam^2$ & 0 & 0 \\
				&  &  && & & $3<\lam^2<24/7$ &  & & & \\
				&  &  &&  & & Stable spiral for &  &  & & \\
				&  &  &&  & & $\lam^2>24/7$ &  &  & & \\
				\hline
				$\rm C_\pm$ &  $\pm 1$ & 0 & 0 & 0 & For all $\lambda$ & Saddle point  &  0 & 1 & 1  & 1 \\ 
				\hline
				$\rm D_\pm$ & $\lambda/\sqrt{6}$ & $\pm\sqrt{1-\lam^2/6}$ & 0 & 0 & $\lambda^2 < 6$ &Stable node for $\lambda^2 < 3$ & 0 & 1 & $-1+\lambda^2/3$  & $-1+\lambda^2/3$ \\ 
				& && & & & Saddle point for $3< \lambda^2 < 6$ & & &  &\\
				\hline
				\hline
		\end{tabular}}
	\end{center}
	\caption[crit]{\label{crit} Fixed points with their nature of stability and existence conditions are given for the autonomous system (\ref{eq:x})-(\ref{eq:omr}).}
	\label{tab:fp}
\end{table*}

The points $\rm A_\pm$ and $\rm B_\pm$ can represent radiation and matter dominated eras respectively, provided the value of $\lambda$ is sufficiently larger than what is required for the existence of these points. The critical points $\rm C_\pm$ represent a kinetic regime where the kinetic term of the scalar field dominates with $w_{\rm eff}=w_\pi=1$. The points $\rm D_\pm$ can lead to an attractor solution close to de Sitter for small $\lam$. Restricting our attention to a possible future attractor solution,  we are then left with four points ($\rm B_\pm$ and $\rm D_\pm$). Among these, for the points $\rm B_\pm$, neither the scalar field nor ordinary matter fields dominate entirely and there is a scaling solution \cite{Copeland:1997et} where the energy density of the scalar field remains proportional to that of the background fluid, which in this case is baryonic matter. The condition $\lam^2>3$ has to be satisfied for this solution. However, this solution cannot give us late-time acceleration as $w_{\rm eff}=w_\pi=0$ for the points $\rm B_\pm$. Therefore, from observation we can rule out these points. In other words, we can conclude that the a (steep) exponential potential with $\lam^2>3$ cannot give late-time acceleration; rather, it leads to  a scaling solution for which $w_{\rm eff}=w_\pi=0$, similar as the quintessence field with exponential potential \cite{Copeland:1997et}. Note that the same scaling behaviour is also present during the radiation era but it is not an attractor solution. 

So we are left with the option of $\lam^2<3$ {\it i.e.}, the condition required for the critical points $\rm D_\pm$. From Table~\ref{tab:fp}, it is easy to understand that as we lower the values of $\lam$ we get closer to the de Sitter solution ($w_\pi=-1$). But for these points we do not have such scaling behaviour. 

\subsection{Imposing the Swampland constraints}
As described in the introduction, for the EFTs, responsible for such late-time accelerations, to be compatible with quantum gravity, we need to impose the swampland criteria on the cosmological dynamics of our model. For this, we shall primarily have to focus on the dS constraint (S2) since the condition (S1) is automatically satisfied for our case. Moreover, for a potential of the exponential type with our defined parameters, (S2) translates into a condition for the first derivative of the potential and we do not have to worry about the refined version of (S2), involving the Hessian of the potential, since it is not relevant in this case. It is worth pointing out that this is the same condition which has been applied previously in quintessence models as well as for some Horndeski theories for constraining them. Therefore, from now on, we shall further impose that $\lam\approx 1$, for models which do not fall into the swampland. 

This implies that $w_{\rm eff}=w_\pi=-2/3$ at the fixed points $\rm D_\pm$ in the future. (In hindsight, we could have ruled out some of the critical points simply on the basis of the swampland conjecture (S2).) However, this value of $w_{\rm eff}$ is obviously not compatible with current observational results. So we have to fix the initial conditions very precisely so that the current value of the equation of state ($w_\pi$) matches with observational data but does, however, deviate from that value in the future to reach the fixed point value {\it i.e.}, $w_\pi=-2/3$ for $\lam=1$. To achieve this we have to consider the thawing dynamics \cite{Caldwell:2005tm} of the scalar field where the scalar field behaves like a cosmological constant for most of the history of the universe but starts evolving from the recent past and deviates from the value $w_\pi=-1$. It is worth emphasizing at this point that thawing dynamics of the scalar field is extremely sensitive to the initial conditions, just as is the case for the cosmological constant. This is, in fact, also the case in \cite{Heisenberg:2019qxz} where a model of Horndeski higher-derivative self-interactions was considered and the cosmological dynamics were shown to be sensitively dependent on the tuned initial conditions.

For our scenario under consideration, we have one variable more than standard quintessence which gives us some freedom in choosing our initial conditions. In \cite{Hossain:2012qm} it has already been shown that scalar field EoS can have lower values than the quintessence case as we increase the positive initial values of $\ep$ for a constant $\lam$. This is precisely what we shall cleverly employ so as to have the values of $w_\pi$ consistent with both current observation as well as the swampland criteria.

\begin{figure}[h]
	\centering
	\subfigure[Cosmological evolution of $\ep$ for three different initial conditions with $\lam=1$.]{\includegraphics[scale=.7]{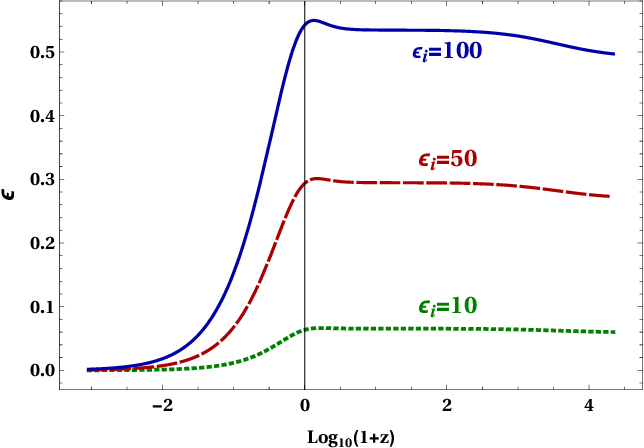}\label{fig:ep}}~~~~~
	\subfigure[Green (dotted), red (dashed) and blue (solid) curves correspond to the cosmological evolution of $w_\pi$ for $\ep_i=0,\; 10,\; 20$ respectively with $\lam=1$.]{\includegraphics[scale=.7]{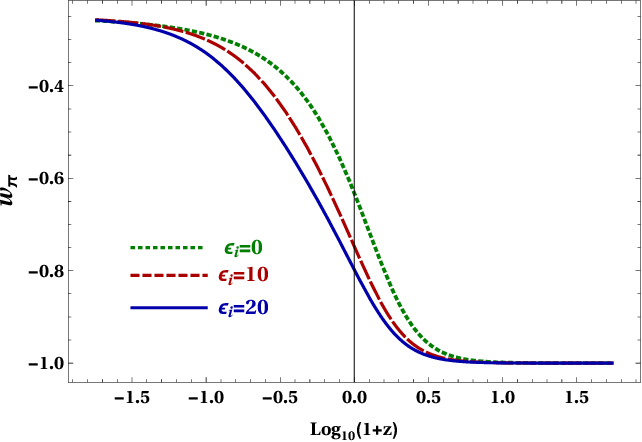}\label{fig:eos}}\\
	\subfigure[Cosmological evolution of the fractional energy densities are shown for $\lam=1$.]{\includegraphics[scale=.7]{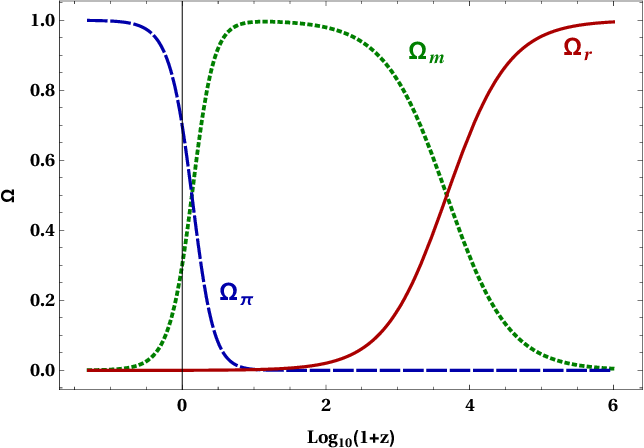}\label{fig:den}}~~~~~
	\subfigure[Cosmological evolution of the dimensionless parameters $x,\; y$ and $\ep$ are shown for $\lam=1$.]{\includegraphics[scale=.7]{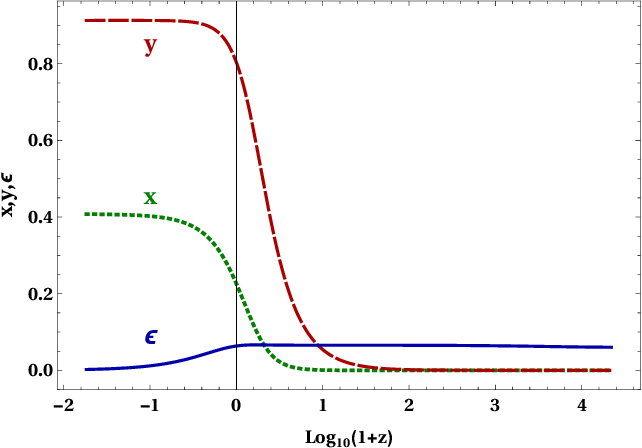}\label{fig:den1}}
	\caption{Evolution of cosmological parameters in our model}
\end{figure}

As briefly mentioned earlier, from Table~\ref{tab:fp} we can see that at all the fixed points are at $\ep=0$. But we are mainly interested about the late time attractor solution and an initial condition dependent cosmological evolution. In other words, we shall fix arbitrary initial conditions so as to reproduce a viable cosmological evolution with the correct amount of radiation, matter and scalar-dominated phases, and shall then study the late-time dynamics. Of course, even if we set arbitrary initial conditions, we should reach one of the points of $\rm D_\pm$ in future where $\ep=0$ which has been shown in Fig.~\ref{fig:ep}. That means in future the dynamics of the model should reduce to that of the quintessence. This has been illustrated in Fig.~\ref{fig:eos} with the evolution of the scalar field EoS where the red (dashed) and blue (solid) curves are approaching the green (dotted) curve which is for $\ep_i=0$ {\it i.e.}, the quintessence case. The cosmological evolution of the fractional energy density is shown in Fig.~\ref{fig:den} which shows the viability of the cosmological dynamics with the chosen initial conditions.

The evolution of the dimensionless quantities $(x, y, \ep)$ is shown in Fig.~\ref{fig:den1}. At the onset, the scalar field is kept nearly frozen due to large Hubble damping which allows us to choose very small values for initial values of $x$ ($x_i$), and keeps $w_\pi \approx -1$. On setting the initial value of $\ep$ \textit{i.e.}, $\epsilon_i$ to zero, it remains zero throughout the cosmic history and we flow along the trajectories of standard quintessence fields to the fixed points at late times. Specifically for $\ep_i=0$, the situation reduces to that of quintessence models with thawing dynamics, for which we have tuned initial value of $y$ ($y_i$) so as to get a viable cosmology. On fixing a suitable $y_i$, the quintessence solution for $\lam\approx 1$ is largely independent of the initial condition $x_i$ \textit{at late times}. On the other hand, in the presence of the higher derivative cubic Galileon term, \textit{i.e.} when there is a non-zero $\ep_i$, different values of $x_i$ can lead to different cosmological histories near $z=0$. The reason for this is that when $\ep_i$ is set to non-zero positive values, an additional frictional term is turned on which is coupled to the $x$ parameter. Finally, it is worth pointing out that only the relative magnitude between the derivative self-interaction term, $\ep_i$ and $x_i$ is what matters for the dynamics.

\section{Current observational bounds}

Imposing experimental bounds from SNeIA, CMB, BAO and $H_0$ data, one can obtain $1\sig,\; 2\sig$ and $3\sigma$ contours for an upper bound on the dark energy EoS as a function of redshift \cite{Scolnic:2017caz}. This was carried out in \cite{Heisenberg:2018yae} for standard quintessence models using the observational constraints on $w_0$ and $w_a$, the parameters of the Chevallier-Polarski-Linder (CPL) parameterization \cite{Chevallier:2000qy,Linder:2002et} of the dark enrgy EoS
\begin{equation}
	w(z)=w_0+w_a\frac{z}{1+z} \, .
\end{equation}
We shall closely follow the analysis of \cite{Heisenberg:2018yae} to put constraints on our model involving higher derivative terms. 

On setting $\ep=0$, one goes back to the quintessence scenario. As is expected analytically for this case, the numerical evolution setting  $\ep_i=0$ reproduces the numerical solution for quintessence. Changing the values of $\ep_i$ changes the evolution history \cite{Hossain:2012qm} and similarly for different $x_i$ (when $\ep_i\neq 0$). Here, we are interested to see the effect of different values of $\ep_i$ and $x_i$ on the evolution history and comparing the latter with observational bounds while respecting the swampland criteria, in particular the dS constraint. Thus, we shall consider $\ep_i$ and $x_i$ as parameters while we shall fix $y_i$ and $\Om_\r$ ($\Om_{\r i}$) to $y_i=5\times 10^{-12}$ and $\Om_{\r i}=0.999$.

\begin{figure}[h]
	\centering
	\subfigure[Green (dotted), red (dashed) and blue (dot dashed) curves correspond to $\ep_i=0,\; 10,\; 100$ respectively. The solid lines represent the $1\sig,\; 2\sig$ and $3\sigma$ contours from bottom to top respectively for the dark energy EoS considering CPL parameterization.]{\includegraphics[scale=.7]{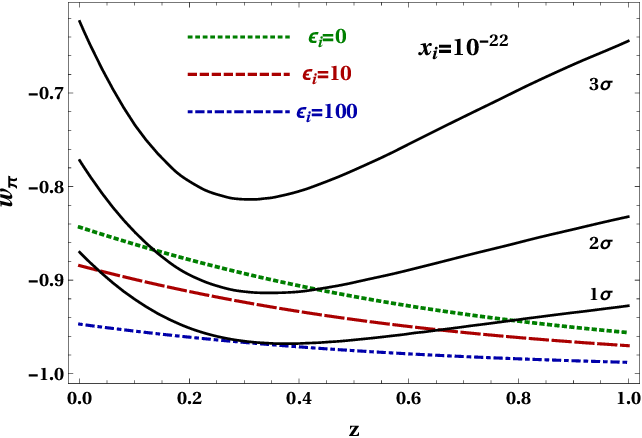}\label{fig:wpi}}~~~~~
	\subfigure[Green (dotted), red (dashed) and blue (dot dashed) curves correspond to $x_i=10^{-22},\; 10^{-23},\; 10^{-24}$ respectively. The solid lines represent the $1\sig,\; 2\sig$ and $3\sigma$ contours from bottom to top respectively for the dark energy EoS considering CPL parameterization.]{\includegraphics[scale=.7]{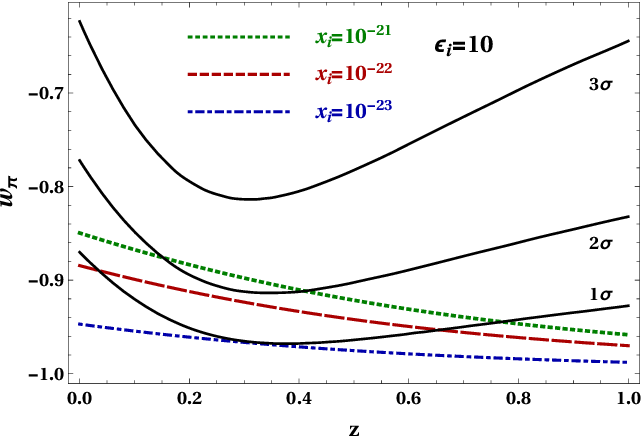}\label{fig:wpi2}}
	\caption{Observational bounds on the reconstructed Galileon EoS as a function of redshift} 
\end{figure}

Fig.~\ref{fig:wpi} shows the constraints on the model from observations for different initial values of $\ep$ by fixing $x_i$. The solid black  lines represent $1\sig,\; 2\sig$ and $3\sigma$ upper bounds from bottom to top respectively on the dark energy EoS. Throughout this analyses, we have fixed the slope $\lam$ to 1 to respect the dS constraint (S2).

The green (dashed) line is for $\ep_i=0$ {\it i.e.}, the quintessence case. As can be seen clearly from the Fig.~\ref{fig:wpi}, this case is in conflict with current observation at $2\sig$ level. As we increase the values of $\ep_i$, the EoS gets closer to $-1$ making it more viable with  observation. Also for a given value of $\ep_i$, we see that one gets the same behaviour if we lower the values of $x_i$ which we have shown in Fig.~\ref{fig:wpi2}. In other words, Fig.~\ref{fig:wpi} and \ref{fig:wpi2} show that for $\lam\sim \mathcal{O}(1)$, even though the quintessence model (with an exponential potential, which is the least constrained case) can be ruled out at the 2$\sig$ level, the inclusion of the cubic Galileon term can make the cosmology viable with both observations and the swampland criteria.

A natural question to address would be that, if the cubic term is made larger (and comparable to the standard kinetic term) to begin with, how can we ignore other higher derivative and higher curvature terms? We have already explained earlier how observational constraints naturally lead us to keep only the cubic term for Galileon theories. Moreover, our aim is not to demonstrate what is the most general modified gravity theory, allowing for all possible higher curvature terms, which is compatible with the Swampland. Instead, our limited goal is to show that there exists viable models that fall within the sub-class of Horndeski Lagrangian, which can be made to satisfy the Swampland criteria. Thus, our goal is not to construct the most general low-energy effective theory keeping all the possible terms but rather to focus on a specific model consistent with observations, and its viability for UV-completeness a la the Swampland.

Another important question which arises is how does our results match with those of \cite{Heisenberg:2019qxz}, which seem to suggest that Horndeski theories are typically in the Swampland? Firstly, we consider only a cubic higher derivative self-interaction term and more importantly, we take the limit where there is no non-minimal coupling in the theory. It is not straightforward to take this limit in the model studied in \cite{Heisenberg:2019qxz} due to the fact that the non-minimal coupling parameter has to be $\geq 10^{-3}$ to comply with solar system constraints \cite{Kase:2015zva}. Therefore, our model is indeed able to scan parts of the phase space of Horndeski theory which is truly unavailable in the system studied in \cite{Heisenberg:2019qxz}, therefore showing the possibility that there exists models of dark energy which are compatible with current observations, even after strongly ($\lam = 1$) imposing the Swampland constraints.

\section{Looking Ahead: More general models}
The swampland conjectures have paved a way towards constraining cosmological models of both the early and late universe. If true, these conjectures tell us how complicated UV-completions of GR can leave its imprint on the low-energy effective theory. Specifically, in the context of the dS constraint, these conjectures suggest that there is a naturalness condition such that the slope of the potential of any effective scalar should be related to its potential through an order one number. In other words, (S2) gives us a way to explain not only why the current value of the potential is small, but also why its slope is equally small. However, current models of quintessence leads to bounds on $c$ which can lead to significant tension with this conjecture, thanks to experimental data coming from current as well as near-future experiments. In this work, we showed through an explicit example, how some of these tensions can be relieved in going to a model beyond quintessence, namely by adding the additional higher derivative cubic Galileon term to the action.

The specific Galileon model we have considered here is still viable after multi-messenger gravitational wave discovery as well as with other current observational bounds even after imposing the swampland conjectures. Moreover, as noted in \cite{Heisenberg:2019qxz}, large portions of the parameter space of Horndeski theories can be ruled out by current and near-future data, if the Swampland constraints are taken into account, thereby somehow singling out models of this kind as special. However, it would be more interesting to see what the swampland conjecture implies for models of modified gravity, aimed at explaining late-time acceleration, in a more systematic manner \cite{Heisenberg:2018vsk}. We plan to do this in future by going to more general models beyond-Horndeski and applying the swampland conjectures to them. In this way, we shall be able to constrain these models through a theoretical consistency requirement in addition to observable data. 

\vskip15pt

\noindent\textit{Acknowledgments:}   We are grateful to Robert Brandenberger for comments on an earlier version of this paper. This research was supported in part by the Ministry of Science, ICT \& Future Planning, Gyeongsangbuk-do and Pohang City and the National Research Foundation of Korea (Grant No.: 2018R1D1A1B07049126).

\end{document}